\documentclass{article}

\usepackage{arxiv}

\usepackage[utf8]{inputenc} 
\usepackage[T1]{fontenc}    
\usepackage{hyperref}       
\usepackage{url}            
\usepackage{booktabs}       
\usepackage{amsfonts}       
\usepackage{nicefrac}       
\usepackage{microtype}      
\usepackage{lipsum}
\usepackage{graphicx} 
\usepackage{upgreek} 

\title{Variable Curvature Displays: Optical Designs and Applications for VR/AR/MR Headsets}

\author{Eduard Muslimov $^{1,2,}$*, Thibault Behaghel $^{1}$, Emmanuel Hugot $^{1}$, Kelly Joaquina $^{1}$ and Ilya~Guskov $^{2}$\\
  \texttt{eduard.muslimov@lam.fr} \\
}

\begin{document}
\maketitle

\begin{abstract}
In the present paper, we discuss the design of a projection system with curved display and its enhancement by variably adjusting the curvature. We demonstrate that the focal surface curvature varies significantly with a change of the object position and that it can easily be computed with the Seidel aberration theory. Using this analytically derived curvature value as the starting point, we optimise a refocusable projection system with $90^\circ$ field of view and $F/\#=6.2$. It is demonstrated that such a system can provide stable image quality and illumination when refocusing from infinity to 1.5~m. The gain in spatial resolution is as high as 1.54 times with respect to a flat focal surface. Furthermore, we prove that a silicon die can be curved to the required shape with a safety factor of 4.3 in terms of the mechanical stress. Finally, it is shown that the developed system can be used in a virtual reality headset providing high resolution, low distortion and a flexible focusing mode.
\end{abstract}


\section{Introduction}
The aberration of field curvature and the means to correct it have been considered in optical design studies since the works by Joseph Petzval. During the last decade, it was practically demonstrated that the field curvature of an imaging system can be directly corrected by using a curved image sensor. Several  techniques to produce devices based on curved CMOS (Complementary Metall-Oxide--Semiconductor) sensors were developed together with theoretical studies on the design and analysis of optical system using these detectors \cite{Swain, Rim, Dinyari,Iwert, Hugot, Gaschet}. It was shown that such systems can provide a better image quality with fewer optical elements, and therefore reduced volume and mass. These~systems also have advantages in terms of the image distortion and the illumination uniformity \cite{Rim, Hugot}.

Recently, it has been demonstrated that the approach used for the production of curved sensors \cite{Lombardo} can be applied to create curved microdisplays thanks to the similarity between CMOS detectors and OLED (Organic Light-Emitting Diode) displays \cite{Maindron}. This latest development opens new prospects for the creation of optical systems with curved focal surfaces. Now,  the potential applications expand beyond the frames of ordinary imaging optics to projection systems. The latter area includes emerging topics such as virtual/augmented reality (VR/AR) and head-mounted displays (HMDs). The use of curved displays helps to significantly increase key parameters of the VR/AR devices such as field of view (FoV) and resolution without an increase of their mass and volume, as   shown by \cite{Han}. Currently, the simultaneous enhancement of these parameters represents one of the challenges in the field, which urges designers to use complex solutions with extreme aspheres, FoV tiling, etc.  \cite{Narasimhan,Wartenberg}. In addition, the application of curved displays facilitates the exploration of new design options such as zoom projection systems. However, designing such a system requires completing some initial tasks, which include the choice of starting configuration and optimisation strategy, as well as modelling of the display curving process.   

In the present paper, we consider the approach to optical design with curved display and its extension for a zoom system. The paper is organised as followed. Section~\ref{section2} presents the theoretical background of the optical design with a variable curvature focal surface. In Section~\ref{section3}, we present the design and optimisation algorithm. Section~\ref{section4} provides an optical design example with its performance analysis. In Section~\ref{section5}, we give estimations of the design feasibility in terms of the display curving technology. Lastly, Section~\ref{section6} contains simulations demonstrating a possible application of the design for VR.  


\section{Theoretical Background}
\label{section2}

\begin{figure}[h]
\centering
\includegraphics[width=10cm]{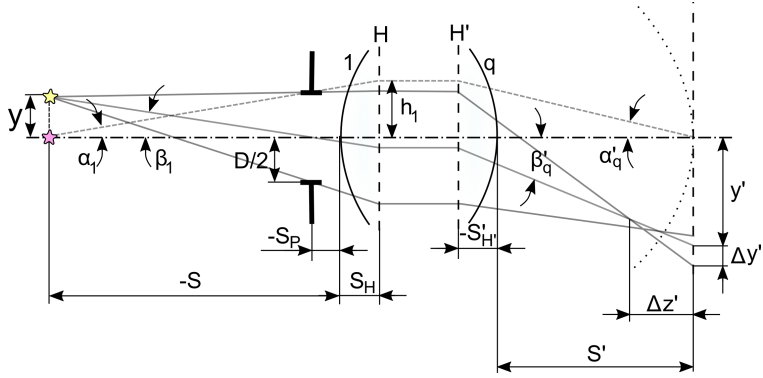}
\caption{
\label{fig:VRtheory}
General raytracing scheme used to derive the field curvature.}
\end{figure}

   To obtain a good starting approximation for the optical design and to demonstrate the field curvature dependence on the system configuration, we consider Seidel aberration theory \cite{{Gross}}. The~general layout of a projection lens is shown in Figure~\ref{fig:VRtheory}. Note  that hereafter we assume that the raytracing procedure is fully reversible. The figure shows tracing of two paraxial rays used for definition of the Seidel sums and the corresponding notation. 
   
The image surface shape is determined by two Seidel sums corresponding to astigmatism and field curvature:
\begin{equation}
\label{eq:S3}
S_{III}=\sum_{i=1}^{q}h_iP_i\big(\frac{\beta_{i+1}-\beta_i}{\alpha_{i+1}-\alpha_i}\big)^2,
\end{equation}
\begin{equation}
\label{eq:S4}
S_{IV}=-\sum_{i=1}^{q}\frac{1}{h_i}\frac{\alpha_{i+1}n_{i+1}-\alpha_in_{i}}{n_i\alpha_i},
\end{equation}
 where
\begin{equation}
\label{eq:P}
P_i=\bigg(\frac{\alpha_{i+1}-\alpha_i}{1/n_{i+1}-1/n_i}\bigg)^2\bigg(\frac{\alpha_{i+1}}{n_{i+1}}-\frac{\alpha_{i}}{n_{i}}\bigg),
\end{equation}
where $n_i$ is the index of refraction after the $i${th} surface. 

In addition, in the aberration equations, the Lagrange invariant is used in the following form:
\begin{equation}
\label{eq:I}
I=-n_1\alpha_1(S-S_P)\beta_1,
\end{equation}

Then,  the longitudinal aberration can be derived for the rays propagating in tangential and sagittal~planes:
\begin{equation}
\label{eq:dzt}
\Delta z_t = \frac{y^2 B}{2n'_q(S-S_P)\alpha_1\alpha'_q \beta_1^2}\big(3S_{III}+I^2S_{IV}\big),
\end{equation}
\begin{equation}
\label{eq:dzs}
\Delta z_s = \frac{y^2 B}{2n'_q(S-S_P)\alpha_1\alpha'_q \beta_1^2}\big(S_{III}+I^2S_{IV}\big),
\end{equation}
where $B$ is the linear magnification.

The corresponding radii of curvature are:
\begin{equation}
\label{eq:rad}
R_{t(s)}=-\frac{\Delta z_{t(s)}^2+y'^2}{2\Delta z_{t(s)}}.
\end{equation}

One can see that the image surface curvature changes with  the change of  both distance to the object $S$ and   magnification $B$.    To demonstrate this effect and to estimate the accuracy of this theoretical approximation, we perform a series of computations for a simple singlet lens \cite{Murakami}. The~lens is scaled to have a focal length $f'=36$~mm and $F/\#=12.9$. The linear FoV is $\pm$12.7~mm. Since Seidel theory works for monochromatic aberrations, we consider a single wavelength $\lambda=780$~nm, which is used in the original patent. Three positions of the objects were considered: infinity, 5 m and 1.5~m. For~each computation, the image surface curvature was calculated with Equation~(\ref{eq:rad}). For~comparison, it was also computed from the marginal ray intersection coordinate and by numerical optimisation of the root mean square (RMS) radius in the spot diagram. Finally, a calculation with a flat image surface in the nominal focus position was used to compare with. For each calculation, the lens ray aberrations were computed. The results are summarised in Table~\ref{tab:demo}.

\begin{table}[h]
\caption{Field curvature computations for demonstrative example (FoV center/corner).}
\label{tab:demo}
\centering
\begin{tabular}{clcccc}

\toprule
\textbf{Distance,~m} & \textbf{Criterion}       & \textbf{R,~m} & \textbf{Spot X,}~\boldmath{$\upmu$}\textbf{m} & \textbf{Spot Y,}~\boldmath{$\upmu$}\textbf{m} &  \textbf{Spot RMS,}~\boldmath{$\upmu$}\textbf{m} \\ 
\midrule
1.5           & Seidel Sum Tang. & 128.55 &   12.7/46.2     &     12.7/27.9     &   9.0/27.5          \\
{}            & Seidel Sum Sag.  &  75.12 &  26.6/25.4      & 26.6/43.2        &    15.2/28.7   \\
{}            & Marginal ray    &     151.87 &   9.8/50.6       &   9.8/25.1       &  8.8/28.4 \\
{}            & RMS              & 117.44 &   14.6/43.4      &     14.6/29.9    &  9.4/27.2       \\
{}            & None             & inf.   &    21.1/74.6     &   21.1/23.0       &  15.9/38.0     \\
\midrule
5             & Seidel Sum Tang. & 129.50 &    12.1/46.4      &   12.1/ 26.6    &    9.0/27.2          \\
{}            & Seidel Sum Sag.  & 74.23  &    26.3/25.1     &    26.3/41.9    &    15.1/27.9     \\
{}            & Marginal ray     & 140.67 &   10.6/48.7     &     10.6/25.    &   8.9/27.7    \\
{}            & RMS              & 113.59 &   14.8/42.5      &    14.8/29.2     &  9.5/26.7       \\
{}            & None            & inf.   &   21.5/74.3       &    21.5/24.1   & 16.0/38.0     \\
\midrule
inf.          & Seidel Sum Tang. & 129.89 &    11.9/46.5    & 11.9/26.1         &     9.0/27.2   \\
{}            & Seidel Sum Sag.  & 73.87  &     26.2/25.0    &     26.2/41.4    &    15.0/28.0     \\
{}            & Marginal ray     & 136.53 &      10.9/47.8   & 10.9/25.1        &    9.0/27.4       \\
{}            & RMS              & 112.08 &      14.9/42.1   &    14.9/29.0     &   9.6/26.5        \\
{}            & None            & inf.   &      21.6/74.1   &  21.6/24.4       &  16.2/38.0      \\

\bottomrule

\end{tabular}
\end{table}

This simple example clearly shows three important points. Firstly, the image quality can be significantly improved by using a curved image surface. Secondly, the curvature changes notably with a change in object position. Thirdly, the value computed with the Seidel sum for rays in the tangential plane provides a good approximation for the curvature and can be computed analytically for any optical system.
 
\section{Design Algorithm}
\label{section3}

\begin{figure}[h]
\centering
\includegraphics[width=10 cm]{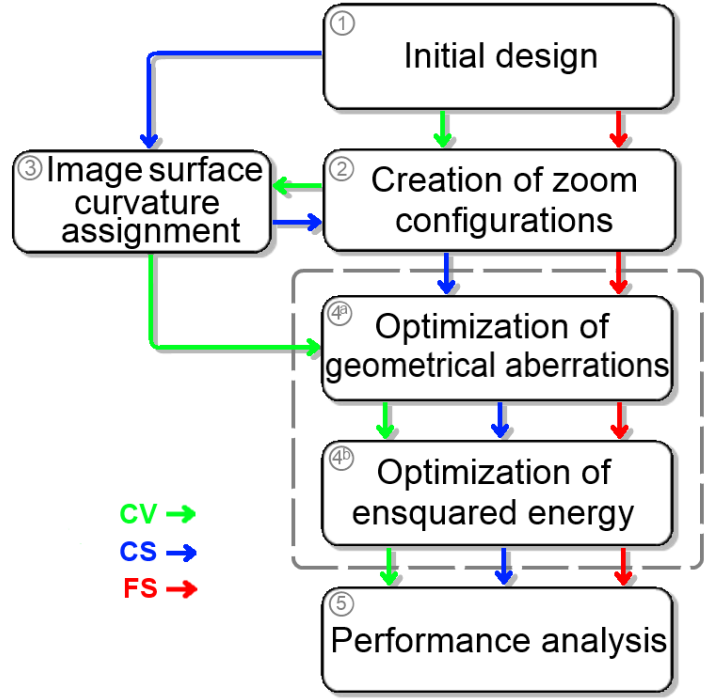}
\caption{
\label{fig:VRalgorithm}
Design algorithm block diagram. The abbreviations indicate the detector configuration: ``FS'', flat, static; ``CS'', curved, static; ``CV'',   curved, variable.}
\end{figure}

 The analytical procedure described above is used for an initial assignment of the image surface curvature in different configurations. This idea underlies the proposed design and optimisation algorithm. The design algorithm is presented as a simplified block diagram 
 in Figure~\ref{fig:VRalgorithm}. If the application implies zooming or refocusing of the optical system, after choice of the initial optical design \textit{(1)}, the design is split into a number of zoom configurations \textit{(2)}, having, e.g., different object positions. At the next step, the initial focal surface curvature for each configuration is computed through Equations~(\ref{eq:S3})--(\ref{eq:rad}) \textit{(3)}. Afterwards, the optical design is optimised for each configuration. It is practically useful to perform two optimisation loops with different quality criteria, e.g., geometrical aberrations \textit{(4a)} and ensquared energy \textit{(4b)}. In both cases, the variables include all the radii, thicknesses and the aspheric coefficients of the optics. The boundary conditions include maintaining the image and object size as well as limitations on the center and edge thicknesses. At the first stage \textit{(4a)}, the merit function is based on the root mean square radii in the spot diagram used with equal weight coefficients. At the second stage \textit{(4b)}, we use  the fraction of energy ensquared in one pixel and the weight coefficient for the edge of the FoV  is doubled. The~obtained optical system is subject to a detailed performance analysis \textit{(5)}, which includes definition of the energy concentration and some secondary indicators such as the distortion and relative illumination of the image. The main path assumes a varying image surface curvature separately for each zoom position and is marked as \textbf{``CV''} for ``curved, variable''. For comparison, we also consider the cases of static curvature ( \textbf{``CS''} for ``curved, static''), when the same value is assigned to all configurations, and flat surface (\textbf{``FS"} for ``flat, static'').   

In the following  section, we apply this algorithm to design a refocusable lens suitable for VR~applications.

\section{Optical Design Example}
\label{section4}
Targeting VR devices as the primary application, and considering other solutions \cite{Deng, Koneva} in the field, we   chose  the following initial data for the optical design example. The pupil diameter is 6~mm and the distance between the pupil and the first lens is $\geq 18$~mm. The angular field of view is 90$^{\circ}$, and the real image size is $50.8$~mm. The system is designed for the visible domain with the standard \textit{d},\,\textit{F} and \textit{C} reference wavelengths. It should be noted here  that Equations~(\ref{eq:S3})--(\ref{eq:rad}) provide the field curvature correction only around the central wavelength. They are used to find the best starting approximation and then the correction for the entire working range is provided by numerical optimisation, as   shown in Figure~\ref{fig:VRalgorithm}. Further, we consider the three design strategies, explained in the previous section and each time apply three configurations with the object at infinity, 5~m and 1.5~m.

\subsection{Design Description}
\begin{figure}[h]
\centering
\includegraphics[width=10 cm]{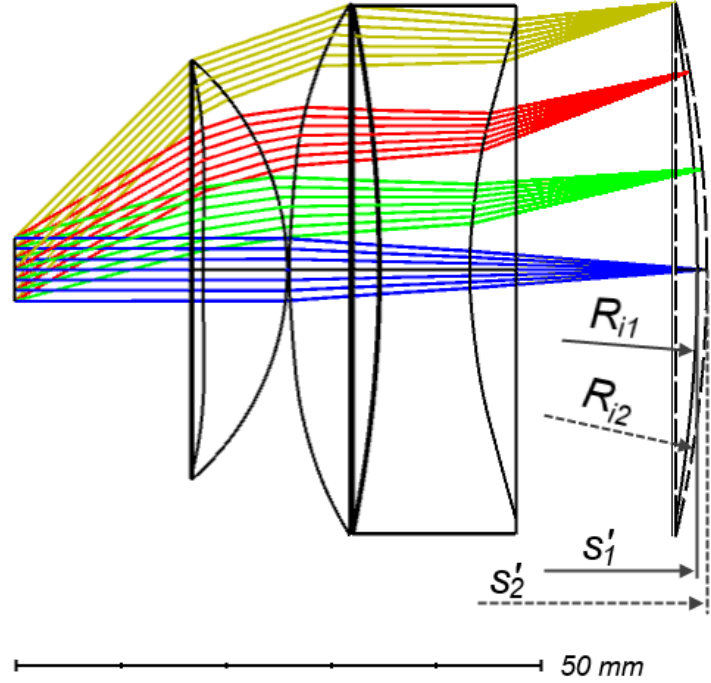}
\caption{
\label{fig:vrlayout}
Overview of the optical design illustrating the variable field curvature concept.}
\end{figure}

A known triplet design \cite{Zobel} was used as the starting point. It was scaled and optimised to match the above given parameters. An overview of the resulting \textbf{``CV''} design is shown in Figure~\ref{fig:vrlayout}. Note that this is an example design and some additional technological optimisation is required for its practical implementation.

The figure clearly shows the difference in the image plane position and curvature between the configuration with object at 1.5~m (``1'' index) and at infinity (``2'' index).
The design contains three Q-type aspheres at the first, third and sixth surfaces with six aspheric coefficients each. The maximum surface slope deviation is 0.163, 0.147 and 0.107, respectively.   

\subsection{Performance Analysis}

\begin{figure}[h]
\centering
\includegraphics[width=1.0\textwidth]{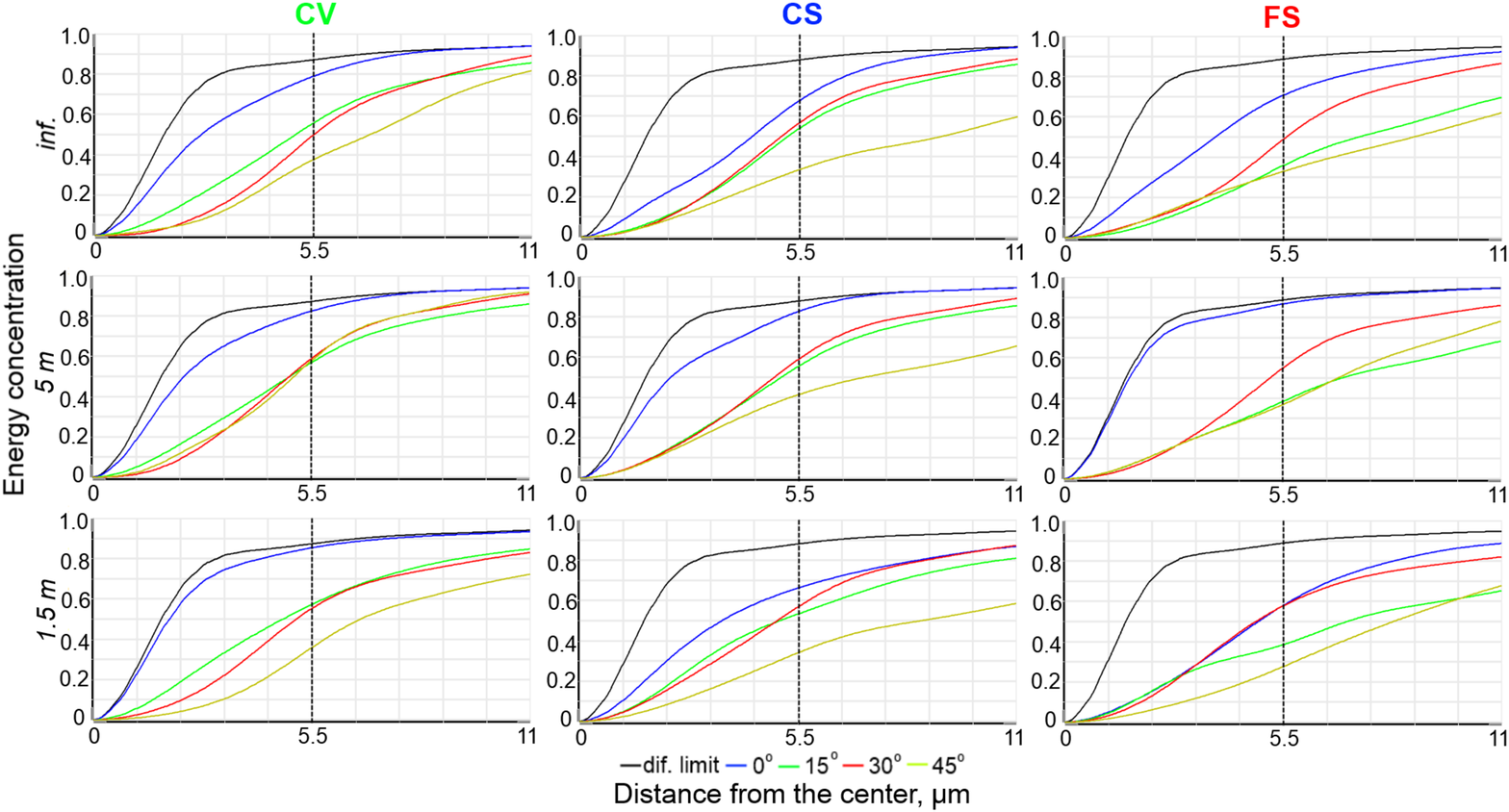}
\caption{
\label{fig:vrenergy}
Energy ensquared in one (11$\upmu$m $\times$ 11 $\upmu$m) and four (22$\upmu$m $\times$ 22 $\upmu$m) pixels} for different designs and different object positions.
\end{figure}

To demonstrate the achieved image quality, we analysed the energy concentration in the point image. The reference element is $11\times11$ $\upmu$m, which corresponds to a typical display pixel \mbox{size \cite{Wartenberg, Richter}}. Additionally, the energy concentration  was compared for $22\times22$ $\upmu$m (i.e., $2\times2$ pixels) area. The~ensquared energy plots for all the configurations are shown in Figure~\ref{fig:vrenergy}.

The grid of subplots clearly demonstrates the advantages of introducing the image surface curvature and being able to variably adjust it. For example, the static curved (``CS'') image surface allows   increasing  the energy concentration for the FoV corner by factor of 1.22. The variable curvature (``CV'') solution makes the values higher by a factor of 1.54 and allows   maintaining  them with a large refocusing.   

We may note here that some secondary performance targets were not included directly into the merit function during the optimisation, yet some gain was obtained. For instance, the \textbf{``CS''} demonstrates significant advantage in the relative illumination and the \textbf{``CV''} configuration shows a good illumination stability with refocusing.  At the same time, the image distortion for these configurations appears to be slightly higher than that for the \textbf{``FS''}. However, the distortion values remain low for all   configurations. The exact numbers are summarised  in Table~\ref{tab:aux}.

\begin{table}[h]
\caption{Auxiliary imaging performance indicators.}
\label{tab:aux}
\centering
\begin{tabular}{ccc}
\toprule
\textbf{Distance,~m} 
& \textbf{Distortion,~$\%$}	& \textbf{Min. rel.illum.} \\
\midrule
	& 	\textbf{CV}  & \\
\midrule
1.5		& 13.2	&	62.0 \\
5		& 13.7	&	60.6 \\
inf.		& 13.9	&	63.6 \\
\midrule
	& 	\textbf{CS}   & \\
\midrule
1.5		& 12.1	&	79.0 \\
5		& 12.1	&	78.3 \\
inf.		& 12.1	& 83.4	 \\
\midrule
	& 	\textbf{FS}   & \\
\midrule
1.5		& 9.2	&	57.8 \\
5		& 9.3	&	59.0 \\
inf.		& 9.3	&	69.4 \\
\bottomrule
\end{tabular}
\end{table}

\section{Variable Curvature Display Feasibility}
\label{section5}
The image surface curvature in the designed lens varies from $-110$  to $-138.7$~mm when refocusing from   infinity to $1.5$~m. The obtained ratio of the display size and curvature is close to the current limit of the curving technology \cite{Jahn}, defined by the die breakage limit. Thus, additional modelling was performed in order to demonstrate the design feasibility.

\begin{figure}[h]
\centering
\includegraphics[width=9 cm]{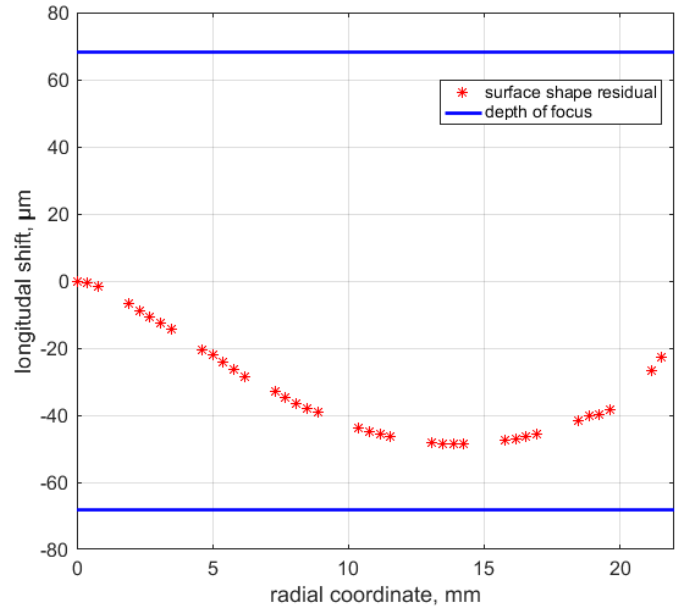}
\caption{
\label{fig:vrdepth}
Difference between the desired display shape and the FEA results compared to the depth of~focus.}
\end{figure}

We estimated the feasibility of variable curvature displays parameters relying on the theoretical basement described in \cite{Lemaitre} and using finite element analysis (FEA) according to the method presented in \cite{Gaschet2, Muslimov}. From this study, to reach the required radii of curvature prescribed from the optical design above, we calculated that the root mean square surface error we can reach is lower than 20~$\upmu$m. The~difference between the desired and actual surface shape along the diagonal is shown in Figure~\ref{fig:vrdepth}. It is compared to the depth of focus, which equals $\pm 68.2~\upmu$m for the lens with $F/\#=6.2$ and $11 \times 11~\upmu$m  pixel. Thus, the calculated display surface shape can be reproduced with a satisfactory precision, although it may be improved by means of a proper mechanical design. 

The corresponding amount of  von Mises Stress generated in the display is lower than 130~MPa, while the breakage limit of silicon is at least $560$~MPa \cite{Coletti}. Thus, the safety factor is about $4.3^\times$. 

In addition, we performed a simple test to estimate the required precision and stability of the curving force. We computed the tolerances on the design parameters, including the display shape. Assuming that a $10\%$  increase of the aberrations is acceptable, we found the tolerance on the display radius equal to $\pm 1.2$~mm. It corresponds to the precision of the curving force within $\pm 1.1\%$. Thus, the display actuator should have a high accuracy and the display should be well isolated from all   other external loads. The latter point should be accounted for in the optomechanical design, which is out of the subject of this study.

This indicates that the proposed design with variable focal plane curvature can be implemented with the known curving technique. However, we must note that this modelling is preliminary and it is intended only  for estimation of the safety factor and the precision requirements. Development of an actual curving setup would require some advanced simulations and additional studies, including nonlinear effects and consideration of technological limitations.
\section{Prospective Application}
\label{section6}
\begin{figure}[h]
\centering
\includegraphics[width=8 cm]{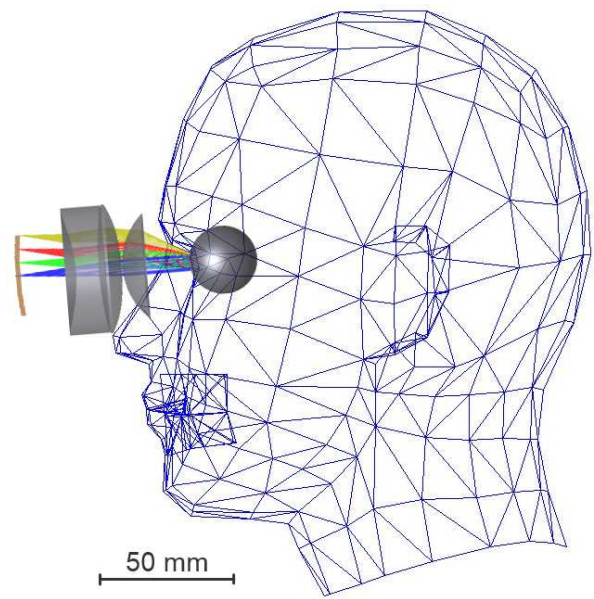}
\caption{
\label{fig:VRface}
Model of the VR projection system setup on a human head.}
\end{figure}

The primary application of the proposed solution is VR/AR helmets. A possibility to project the image at a desired position in depth without a change in the image quality would open new prospects in preparation of the content and interaction with the viewer, especially in VR movies.
Figure~\ref{fig:VRface} shows the optical system model overlapped on the human head model and demonstrates the dimensions and approximate packaging of the helmet. Here, we note that the display used in the design and simulations has a diagonal of 2$^{\prime\prime}$ with the sides ratio of 4:3. The sizes of currently available OLED displays are typically smaller \cite{Haas,Ghosh,Buljan}, but the displays made with this technology \cite{Tsujimura} may reach larger sizes in the nearest future. 

\begin{figure}[h]
\centering
\includegraphics[width=0.9\textwidth]{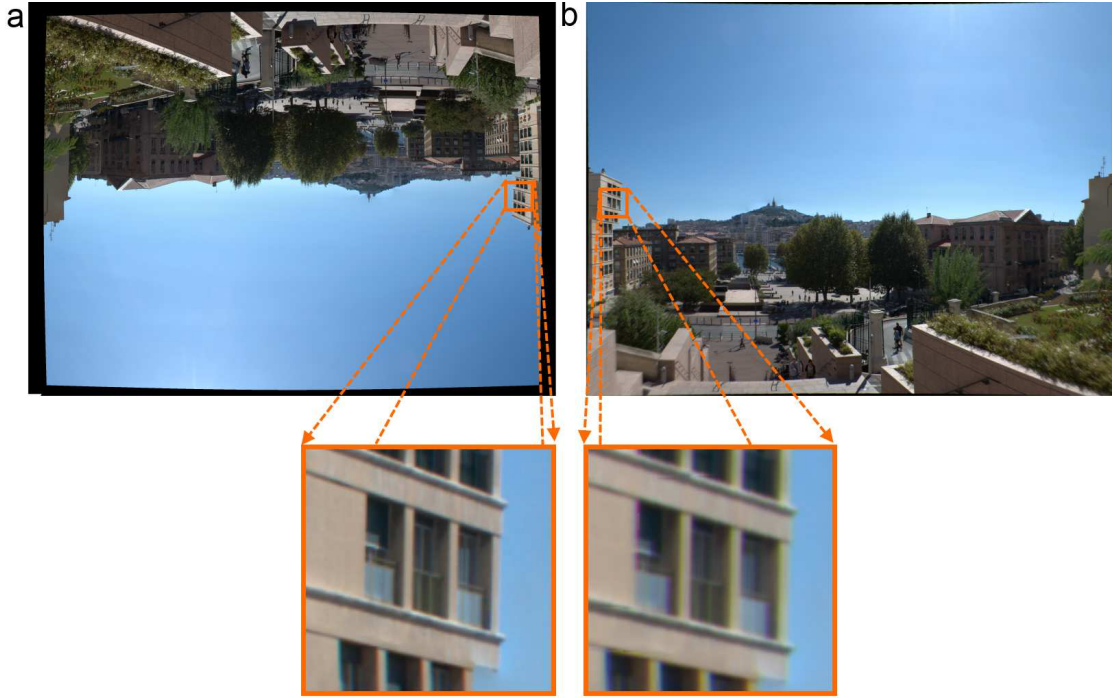}
\caption{
\label{fig:VRsim}
Simulation of the image obtained with the VR system: (\textbf{a})  pre-distorted source image; and (\textbf{b})  simulation of the projected image.}
\end{figure}

 To assess the imaging performance, we performed a simulation in the direct propagation mode (i.e., from the finite distance to infinity). The simulation results are presented in Figure~\ref{fig:VRsim}. One can see that the system provides relatively high image quality even at the FoV edge and the image distortion can be easily removed by pre-processing the projected image. The optical system has a ``pincushion type'' distortion,  thus a distortion with the opposite sign, i.e., ``barrel type'',  should be introduced into the projected image to compensate.

\section{Conclusions}
\label{section7}
We propose  a concept, a design algorithm  and an example design of projection system with a variable curvature display and   show  its performance advantages and physical feasibility. 

The design  algorithm uses Seidel aberration theory to find the initial approximation  and   a multi-configuration two-step numerical optimisation to finalise the design. It was demonstrated that application of this algorithm allows   designing  a projection system with $F/6.2$ aperture and $90^{\circ}$ field of view, which is able to re-focus the projected image from infinity to 1.5~m. The energy concentration in a point source image is higher by a factor of $1.54$ in comparison with a system with flat focal surface. A preliminary modelling shows that a silicon-based display can be curved to the calculated focal surface radius with the breakage safety factor of $4.3^\times$.

The developed optical system, as well as the corresponding design and optimisation approach, can help to improve image quality of the future VR/AR headsets and implement an operational mode with re-focusing. 
However, the prospective application area is much wider and can include, e.g., the design of multimedia projectors and laser scanning systems.


\vspace{6pt} 



\section{Acknowledgements}
This research was funded by the  European Research council through the H2020-ERC-STG-2015-678777 ICARUS program.

The authors would like to thank their colleague Gilles Otten for correction of the manuscript and useful discussions.







\end{document}